\documentclass[twocolumn,pdflatex,sn-mathphys-num]{sn-jnl}

\usepackage{stfloats} 
\usepackage{tabularx}
\usepackage{array} 
\usepackage{mathptmx} %
\usepackage{graphicx}
\usepackage{subcaption}
\usepackage{multirow}%
\usepackage{multicol}
\usepackage{float}
\usepackage{pgfplots}
\usepackage{amsmath,amssymb,amsfonts}%
\usepackage{amsthm}%
\usepackage[title]{appendix}%
\usepackage{xcolor}%
\usepackage{textcomp}%
\usepackage{booktabs}%
\pgfplotsset{compat=1.18}

\makeatletter
\renewcommand\footnoterule{%
  \kern-3pt
  \hrule width \columnwidth
  \kern2.6pt
}
\makeatother

\theoremstyle{thmstyleone}%

\theoremstyle{thmstyletwo}%

\theoremstyle{thmstylethree}%

\begin{document}

\title{
Bootstrapping Audiovisual Speech Recognition in Zero-AV-Resource Scenarios
}

\author*[1,2]{\fnm{Pol} \sur{Buitrago}}\email{pol.buitrago@upc.edu}
\author[1]{\fnm{Javier} \sur{Hernando}}\email{javier.hernando@upc.edu}

\affil[1]{Universitat Politècnica de Catalunya (UPC), Spain}
\affil[2]{Barcelona Supercomputing Center (BSC), Spain}

\abstract{
    Audiovisual speech recognition (AVSR) combines acoustic and visual cues to improve transcription robustness under challenging conditions but remains out of reach for most under-resourced languages due to the lack of labeled video corpora for training. Synthetic visual data have been shown to be an effective augmentation strategy for addressing AV data scarcity. However, a more challenging scenario arises for languages such as Catalan, where no real audiovisual data are available for training.
    
    In this study, we investigate whether AVSR can be bootstrapped in such a zero-AV-resource setting, using synthetic visual data as the sole source of visual supervision. We synthesize over 700 hours of talking-head video and fine-tune a pre-trained AV-HuBERT model. On a manually annotated Catalan benchmark, our model achieves near state-of-the-art (SOTA) performance with much fewer parameters and training data than SOTA ASR systems such as Whisper-large-v3, outperforms an identically trained audio-only baseline, and preserves multimodal advantages under acoustic degradation. Scalable synthetic video thus offers a viable substitute for real recordings in zero-AV-resource AVSR.

}

\keywords{Audiovisual Speech Recognition, Synthetic Data, Zero-AV-Resource, Low-Resource Languages}

\maketitle

\section{Introduction}
\label{sec:intro}

Automatic Speech Recognition (ASR) systems have achieved near-human performance on benchmark datasets, reaching remarkably low error rates under clean, studio-quality audio conditions. However, their performance often degrades substantially in real-world scenarios due to noise, reverberation, channel distortions, or low-bitrate codecs \cite{likhomanenko2021rethinking, mena2026asr,kinoshita2016reverb,kumalija2022asr,hu2024wav2code,besacier2001compression}. Certain phonemic contrasts remain challenging to distinguish from audio alone \cite{files2015visual,miller1955perceptual,mcgurk1976hearing,erber1975auditory,Sumby1954VisualCT}, and in some cases the audio channel may be partially or completely unavailable \cite{battista2025lipreading,Sumby1954VisualCT,marrufoperez2024noise}. 

Audiovisual Speech Recognition (AVSR) addresses these limitations by jointly modeling acoustic signals and visual articulatory cues, particularly lip and facial movements. By combining complementary information from both modalities, AVSR can improve speech recognition under adverse acoustic conditions, enhance intelligibility of degraded speech, and exploit visual cues when acoustic information is unreliable or missing \cite{Sumby1954VisualCT,montesinos2023speech,shi2022avsr,shi2022avhubert,hu2024wav2code}.

State-of-the-art AVSR models typically combine self-supervised pre-training with transformer-based architectures to learn unified representations from large-scale, unlabeled audiovisual (AV) corpora \cite{shi2022avsr,shi2022avhubert,hu2024wav2code,anwar2023muavic}. However, supervised fine-tuning adaptation still requires paired audiovisual data, which remain available for only a limited number of high-resource languages. While many languages have extensive audio-only ASR corpora, the lack of paired video prevents multimodal integration \cite{czyzewski2017audiovisual,anwar2023muavic}.

A natural approach to address this limitation is to generate the missing visual modality from the audio resources that are already available. Recent advances in generative modeling, particularly using Generative Adversarial Networks (GANs) \cite{goodfellow2014gan}, have demonstrated the ability to produce highly realistic synthetic videos \cite{Prajwal2020Wav2Lip,vougioukas2020realistic,li2022speech,rakesh2026talking}. Beyond visual realism, such generative methods can produce lip movements that convey interpretable articulatory cues, which prior studies suggest can benefit both human and machine speech comprehension and visual speech recognition \cite{varano2022speechdriven,yu2024speech,shan2022speech,agarwal2023moocs}.

Although synthetic visual data have been explored for visual speech recognition and lipreading \cite{liu2023synthvsr,yang2024audiovsr}, their use in AVSR remains underexplored. Unlike unimodal lipreading, AVSR requires the effective integration of acoustic and visual representations, making the quality and consistency of synthetic visual cues particularly critical for cross-modal learning. Synthetic videos that provide useful visual information in isolation may therefore fail to benefit AVSR, or even interfere with the learning of acoustic representations that typically provide the dominant source of information. 

In our previous work \cite{buitrago2026improving}, we investigated, to our knowledge for the first time, the use of synthetic visual data for AVSR training and showed that it can be effectively used as a data augmentation strategy, yielding relative WER reductions of up to 16.2\% when real AV data are augmented with synthetic samples. However, this approach still assumes that some real AV data are available for augmentation. For languages with no annotated audiovisual training data at all, a more extreme zero-AV-resource scenario arises. In this setting, only transcribed audio corpora are available in the target language, and any visual supervision during fine-tuning must be generated. Our prior work provided preliminary evidence that synthetic visual supervision alone could support AVSR training in such a setting, but it remained unclear whether such a system could achieve competitive performance relative to strong audio-only ASR models and whether its multimodal advantages would persist under adverse acoustic conditions.

To investigate this setting, we focus on AVSR for Catalan as a case study and fine-tune a pre-trained Audio-Visual Hidden Unit BERT (AV-HuBERT) \cite{shi2022avsr,shi2022avhubert} model using real speech paired exclusively with synthetic visual streams, with no real Catalan audiovisual data used for training. This setup allows us to directly assess whether synthetic visual supervision alone is sufficient to learn useful audiovisual representations in the absence of real visual training data. We then evaluate the resulting system to determine whether it can achieve competitive speech recognition performance while retaining the benefits of multimodal integration under adverse acoustic conditions, and further analyze the properties of the synthetic visual streams underlying this behavior.

To construct the real Catalan audiovisual benchmark used for evaluation, we use the semi-automatic annotation pipeline introduced in our previous work \cite{buitrago2026improving}, which produces manually reviewed audiovisual transcriptions with minimal human effort. Here we extend it with improved documentation and a more accessible user interface, supporting its use as a general-purpose tool for low-resource audiovisual dataset creation.

The aim of this work is to assess whether an effective AVSR system can be trained without real audiovisual training data, using synthetic visual streams as the sole source of visual supervision during fine-tuning. We evaluate whether such a system can match or exceed the performance of state-of-the-art audio-only ASR baselines while retaining the complementary information and robustness benefits provided by the visual modality.

\section{Experimental Setup}
\label{sec:experimental_setup}

This section describes the experimental setup used to investigate AVSR in the zero-AV-resource setting. We first describe the pipeline used to generate synthetic audiovisual training data from audio-only speech and still facial images (Section~\ref{subsec:synthetic_data}). We then introduce the semi-automatic annotation pipeline used to construct the real Catalan audiovisual benchmark for evaluation (Section~\ref{subsec:evaluation}). Finally, we describe the AV-HuBERT model and fine-tuning strategy (Section~\ref{subsec:model_finetune}).

\subsection{Synthetic Audiovisual Data Generation}
\label{subsec:synthetic_data}

\begin{figure*}[t]
\centering
\includegraphics[width=\linewidth]{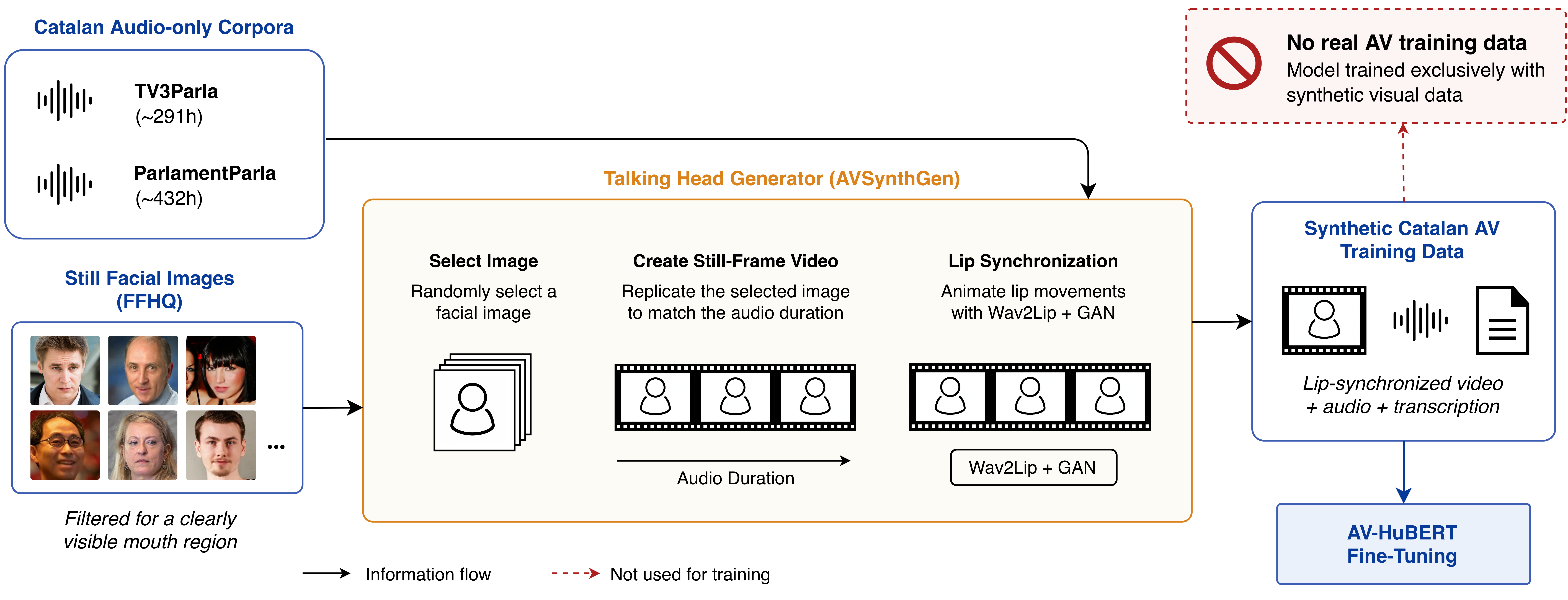}
\caption{\itshape Pipeline for generating synthetic audiovisual data from speech audio and facial images. Still face images are selected and animated with lip movements synchronized to the input speech using a pre-trained Wav2Lip+GAN model \cite{Prajwal2020Wav2Lip,goodfellow2014gan}, producing talking-head videos aligned with the original audio.}
\label{fig:synthetic_pipeline}
\end{figure*}

We generate the synthetic audiovisual training data using \textit{AVSynthGen}, a language-agnostic pipeline introduced in our previous work \cite{buitrago2026improving}. The pipeline converts audio-only speech recordings into lip-synchronized talking-head videos by combining each speech sample with a still facial image. As illustrated in Figure~\ref{fig:synthetic_pipeline}, the process consists of selecting face images with a clearly visible mouth region, creating a video sequence from each image, and animating the mouth movements according to the corresponding speech signal\footnote{Code for the \textit{AVSynthGen} pipeline is available at \url{https://github.com/Pol-Buitrago/SynthAVSR}.}.

\begin{figure}[b]
\centering
\includegraphics[
width=\columnwidth,
trim={0 -20 -30 0},
clip
]{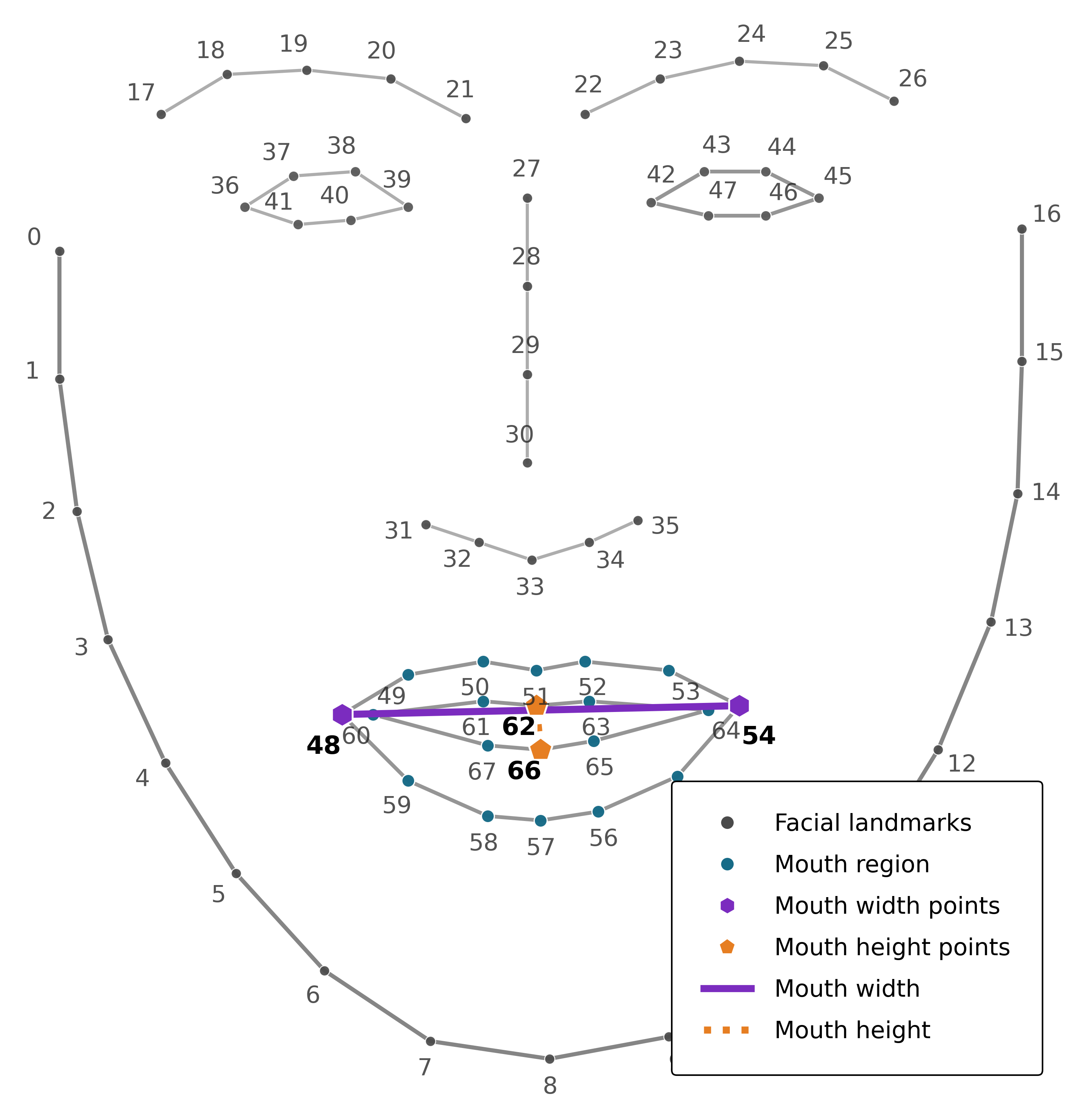}
\caption{\itshape The 68-point facial landmark template used for morphological face filtering. Facial landmarks are extracted using \texttt{dlib} \cite{king2009dlib}.}
\label{fig:landmark_template}
\end{figure}

For this study, the audio used to generate the synthetic videos is sourced from the Catalan TV3Parla \cite{kulebi18_iberspeech} and ParlamentParla \cite{kulebi2022parlamentparla} corpora, comprising approximately 291 and 432 hours of speech, respectively. For the visual component, still facial images are sampled from the Flickr-Faces-HQ dataset (FFHQ) \cite{Karras2019StyleGAN}, which provides diverse identities, poses, and lighting conditions. 

Facial image selection is performed using a landmark-based morphological filtering stage implemented with \texttt{dlib} \cite{king2009dlib}. A frontal face detector is first applied to each image, followed by the extraction of 68 facial landmarks, whose template is shown in Figure~\ref{fig:landmark_template}. The landmarks corresponding to the mouth region are used to assess whether the mouth is sufficiently visible based on simple geometric criteria derived from inter-landmark distances, including mouth width and opening. Images that do not satisfy these visibility criteria are discarded before synthetic video generation.

After filtering to obtain a clearly visible mouth region, each selected image is paired with an audio recording to create a still-frame video in which the same facial image is maintained throughout the duration of the audio. The resulting sequence is then processed by a pre-trained Wav2Lip+GAN model \cite{Prajwal2020Wav2Lip,goodfellow2014gan}, which synthesizes lip movements synchronized with the speech.

The generated videos are paired with their original audio recordings and transcriptions, producing a synthetic audiovisual corpus used to fine-tune the AVSR model. This synthetic corpus provides the complete visual supervision for training, with no real audiovisual data from Catalan used during training. 

After processing all audio samples through the generation pipeline, the resulting synthetic corpus comprises approximately 723 hours of audiovisual data. This entire corpus is used for fine-tuning the AV-HuBERT model for Catalan, providing the complete audiovisual training data available to the model in the zero-AV-resource setting.

\subsection{Semi-Automatic Audiovisual Annotation}
\label{subsec:evaluation}

Although the zero-AV-resource setting investigated in this work relies exclusively on synthetic visual data for training, real audiovisual data remain essential for meaningful evaluation of the resulting system. Evaluating AVSR on synthetic videos would not assess its ability to process real-world visual speech, making an independent benchmark of real audiovisual recordings necessary. 

To construct the real audiovisual benchmark used for evaluation, we extend the semi-automatic audiovisual annotation pipeline presented in our previous work \cite{buitrago2026improving} with improved documentation and a more accessible user interface\footnote{Code is available in the same repository as AVSynthGen, see Section~\ref{subsec:synthetic_data}.}. The pipeline is designed to reduce the manual effort required for annotation while retaining human verification of the final samples. Beyond its use for the Catalan benchmark in this study, this extended pipeline is designed as a general-purpose, language-agnostic tool for efficiently constructing real audiovisual resources in low-resource settings.

\begin{figure*}[t]
	\centering
	\includegraphics[width=\linewidth]{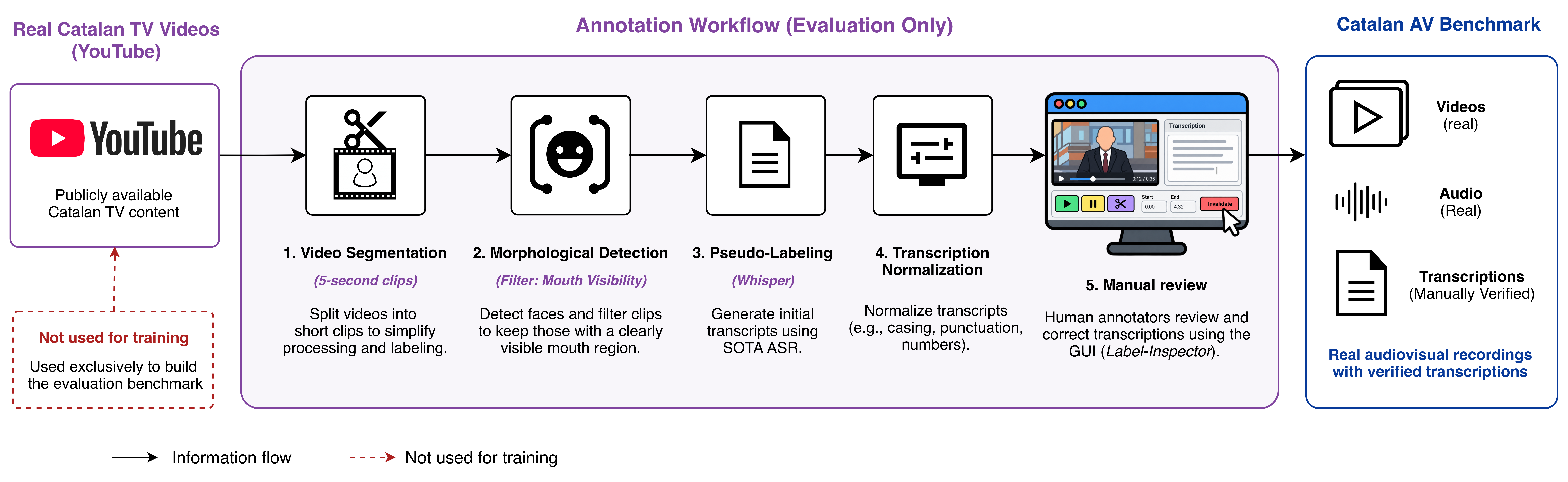}
	\caption{\itshape Semi-automatic pipeline for annotating the Catalan AV benchmark. The process combines segment extraction, morphological filtering, automatic pseudo-labeling \cite{radford2022whisper}, and manual verification and refinement via a custom graphical interface (GUI).}
	\label{fig:annotation_pipeline}
\end{figure*}

As illustrated in Figure~\ref{fig:annotation_pipeline}, the workflow combines automatic video processing, face and mouth detection, speech recognition, transcription normalization, and human verification through a lightweight graphical interface. Starting from publicly available video, the source material is first segmented into short video clips of approximately five seconds, with segment boundaries placed at pauses or silent intervals whenever possible to avoid splitting words across clips. 

A morphological detection stage is then applied to identify segments containing a clearly visible face and mouth region. Facial landmarks are extracted using \texttt{dlib} \cite{king2009dlib}, and simple geometric criteria based on mouth width and opening are used to filter out samples with insufficient mouth visibility for audiovisual speech recognition. For the remaining clips, pseudo-transcriptions are automatically generated using a pre-trained Whisper model \cite{radford2022whisper}. These transcriptions are subsequently normalized to obtain a consistent textual representation.

\begin{figure*}[t]
\centering
\includegraphics[width=0.9\linewidth]{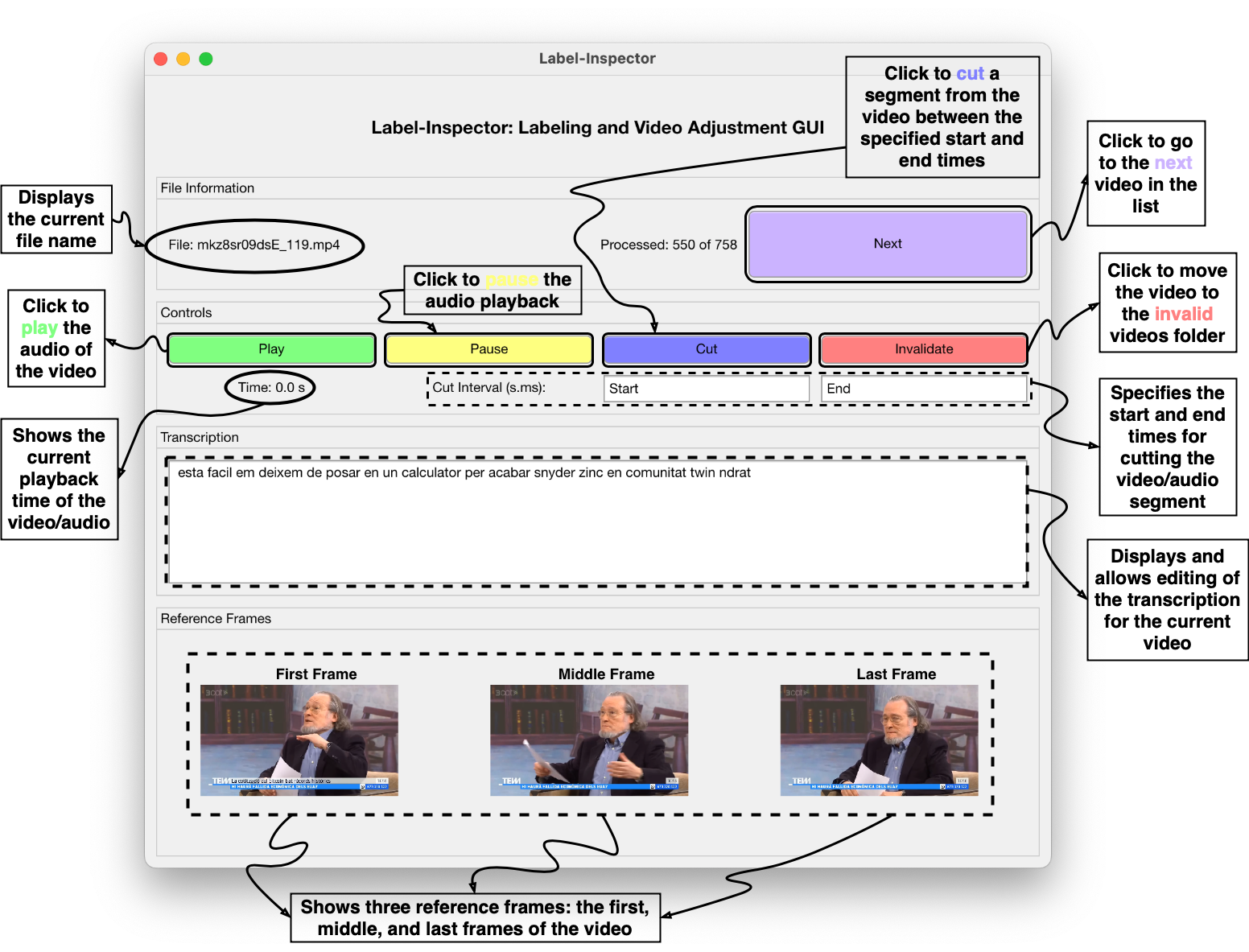}
\caption{\itshape Label-Inspector GUI used for human verification and refinement of automatically generated audiovisual annotations. The interface provides audio and video playback, precise temporal boundary adjustment, clip cutting and invalidation controls, editable transcription, and reference frames from different points of each clip to facilitate rapid inspection of temporal alignment and visual quality.}
\label{fig:label_inspector}
\end{figure*}

The resulting video segments and automatically generated transcriptions are then presented to a human annotator through the \textit{Label-Inspector} graphical user interface (GUI). Figure~\ref{fig:label_inspector} shows the interface used for this stage. The annotator can play back the audio and inspect the automatically generated transcription, allowing the temporal boundaries of each clip to be refined with millisecond-level precision. This enables unwanted portions to be removed, for example when words are truncated or other undesirable artifacts are present. The interface also provides reference frames from different points of the clip, allowing the annotator to inspect the visual content and verify that the mouth is clearly visible and adequately aligned with the speech. Samples with insufficient visual quality or unsuitable mouth visibility can be marked as invalid. Most of this filtering is performed automatically by the preceding morphological face-detection stage, so the human annotator primarily performs targeted verification and refinement of the automatically processed samples. 

This workflow substantially reduces the amount of manual processing required to construct real audiovisual resources, since segmentation, visual filtering, transcription, and normalization are performed automatically before human inspection. In this work, it is applied to publicly available Catalan television footage to construct the evaluation benchmark, yielding 51 minutes and 38 seconds of manually reviewed audiovisual data. These real audiovisual samples are used exclusively for evaluation and are not included in model training. 

More generally, the pipeline provides a practical and scalable approach for creating real audiovisual datasets in low-resource settings, where the cost of manually segmenting, transcribing, and verifying audiovisual recordings can constitute a major bottleneck to dataset construction.

\subsection{Model and Fine-Tuning}
\label{subsec:model_finetune}

We use the AV-HuBERT framework \cite{shi2022avhubert,shi2022avsr} as the backbone for our experiments. AV-HuBERT is an established self-supervised framework for learning joint audiovisual representations and provides publicly available pre-trained checkpoints that can be adapted to downstream AVSR tasks. We initialize the model from the publicly available \textsc{Large} checkpoint pre-trained on unlabeled audiovisual data from the English LRS3~\cite{afouras2018lrs3ted} and VoxCeleb2~\cite{chung2018voxceleb2} datasets\footnote{Referenced checkpoint: \url{https://dl.fbaipublicfiles.com/avhubert/model/lrs3_vox/clean-pretrain/large_vox_iter5.pt}}. Although the pre-training data contain only English speech, prior research on multilingual pre-training has shown significant transfer capabilities to other languages \cite{conneau2020xlmr}, suggesting that similar benefits can be leveraged for AVSR in low-resource scenarios.

For Catalan adaptation, we use an attention-based sequence-to-sequence fine-tuning setup with cross-entropy loss \cite{bahdanau2016endtoend}, as described by \cite{shi2022avhubert}. A randomly initialized 6-layer Transformer decoder is attached to the pre-trained encoder to map its output representations to SentencePiece unigram subword units \cite{kudo2018subword}. The model is optimized using the Adam optimizer \cite{kingma2015adam} with a base learning rate of $1\times10^{-3}$ and a tri-stage learning-rate schedule with warmup and decay. To stabilize adaptation to the target language, the pre-trained encoder is frozen for the first 22,500 updates, after which all encoder parameters are unfrozen and jointly fine-tuned with the decoder.

Model selection is based on validation performance, with the selected model subsequently evaluated on the real Catalan audiovisual benchmark described in Section~\ref{subsec:evaluation}.

\section{Experiments}

In this section, we evaluate the proposed zero-AV-resource AVSR approach for Catalan. We first assess the recognition performance of the model trained using real audio paired exclusively with synthetic visual streams, comparing it with an identically trained audio-only baseline (Section~\ref{subsec:cat_model_performance}). We then compare the resulting AVSR system with strong SOTA audio-only ASR baselines (Section~\ref{subsec:baseline_comparison}) and evaluate its robustness benefits under acoustic degradation (Section~\ref{subsec:noise}). Finally, we analyze the properties of the synthetic visual streams underlying these results (Section~\ref{sec:analysis}).

\subsection{Performance on Zero-AV-Resource Catalan}
\label{subsec:cat_model_performance}

We first evaluate whether synthetic visual streams can provide sufficient visual supervision to train an effective AVSR system in the complete absence of real Catalan audiovisual training data. Using the synthetic audiovisual corpus described in Section~\ref{subsec:synthetic_data}, we fine-tune three modality variants of the same AV-HuBERT architecture using the same data and settings, and evaluate them on the manually annotated real Catalan benchmark described in Section~\ref{subsec:evaluation}. We train an audiovisual (AV) model using both audio and visual modalities, an audio-only (A) model with the visual input masked, and a video-only (V) model with the audio input masked. This controlled comparison allows us to assess both the overall effectiveness of the resulting AVSR system and the extent to which it benefits from the synthetic visual modality.

\begin{table}[t]
    \centering
    \setlength{\tabcolsep}{10pt}
    \resizebox{\linewidth}{!}{%
    \begin{tabular}{@{}lccc@{}}
        \toprule
        \textbf{Model} & \textbf{AV} & \textbf{A} & \textbf{V} \\
        \midrule
        \textbf{Catalan AVSR \textsubscript{(Our model)}} & \textbf{19.6\%} & 23.1\% & 105\% \\
        \bottomrule
    \end{tabular}%
    }
    \caption{WER (\%) on the real Catalan audiovisual benchmark for three separately trained AV-HuBERT variants: audiovisual (AV), audio-only (A), and video-only (V).}
    \label{tab:results_cat}
\end{table}

As shown in Table~\ref{tab:results_cat}, the AV model achieves a WER of 19.6\%, compared with 23.1\% for the audio-only model, corresponding to a relative WER reduction of 15.2\%. Since both models are trained using the same real audio and differ only in the availability of the synthetic visual stream, this improvement indicates that the generated visual cues provide a useful representation for the model to exploit.

The video-only model achieves a WER of 105\%, substantially worse than both multimodal and audio-only configurations. This result indicates that the synthetic visual modality alone does not provide sufficient information for reliable speech transcription. Synthesized visual streams remain less intelligible than real video, consistent with previous studies that train visual speech recognition systems on synthetic video alone \cite{shan2022speech,liu2023synthvsr}, which have similarly reported video-only WERs exceeding 100\%. Nevertheless, their contribution when combined with audio demonstrates that the generated lip movements contain useful articulatory information that the multimodal model can exploit to complement the acoustic signal. The results therefore show that synthetic visual supervision can be effectively integrated with real speech to bootstrap an AVSR system in a zero-AV-resource language.

\subsection{Comparison with Baselines}
\label{subsec:baseline_comparison}

Although our approach has been validated in a zero-AV-resource scenario, it is important to contextualize its performance against strong unimodal baselines. While the previous experiment establishes that the synthetic visual modality provides a benefit over the corresponding audio-only model, the ultimate motivation for developing an AVSR system is to determine whether this multimodal approach can provide competitive performance relative to existing transcription systems.

Since there are no other Catalan-specific AVSR systems available, we compare our model with two versions of Whisper~\cite{radford2022whisper}, a state-of-the-art ASR model: 
(i) Whisper-large, trained on approximately 680k hours of labeled multilingual audio (including 1,883 hours of Catalan), and 
(ii) Whisper-large-v3, trained on 1M hours of weakly labeled plus 4M hours of pseudo-labeled multilingual audio~\cite{openai2023whisperlargev3}. As its per-language breakdown is not disclosed, we take the 1,883 hours of Whisper-large as a conservative lower bound for its Catalan data. Table~\ref{tab:comparison_baselines} summarizes the WER of our Catalan AVSR model and the Whisper variants on our manually annotated benchmark, along with their model characteristics.

\begin{table}[t]
    \centering
    \setlength{\tabcolsep}{7pt}
    \resizebox{\linewidth}{!}{%
    \begin{tabular}{@{}lccc@{}}
        \toprule
        \textbf{Model}& \textbf{WER} & \textbf{Catalan Data} & \textbf{Parameters} \\
        \midrule
        Our model \textsubscript{(AV)} & 19.6\% & 723h & 325M \\
        Whisper-large & 31.4\% & 1,883h & 1,550M \\
        Whisper-large-v3 & 18.3\% & \textgreater1,883h
        & 1,550M \\
        \bottomrule
    \end{tabular}%
    }
    \caption{WER comparison on our manually annotated Catalan AV test set. Results compare our AVSR model with two state-of-the-art Whisper variants.}
    \label{tab:comparison_baselines}
\end{table}

As shown in Table~\ref{tab:comparison_baselines}, both Whisper variants have a substantial advantage, being significantly larger and trained on orders of magnitude more data than our model. Nevertheless, our AVSR model, fine-tuned on only $\approx$723 hours of Catalan audio paired with synthetic visual streams, approaches the performance of Whisper-large-v3 and clearly outperforms Whisper-large, while being much smaller. 

This demonstrates the effectiveness of our approach: even with synthetic visual cues, integrating audiovisual information allows the model to compensate for limited data and smaller capacity, achieving competitive performance under much more constrained conditions.

\subsection{Robustness under Additive Noise}
\label{subsec:noise}

Beyond clean-condition accuracy, a primary motivation for developing an AVSR model is to assess whether multimodal integration (here, acoustic input combined with synthetic visual features) can improve robustness under challenging acoustic conditions. To support this claim, we compare the AV version of our model against Whisper-large and Whisper-large-v3 across a range of additive white Gaussian noise levels. Noise is applied to the audio input at different signal-to-noise ratios (SNRs), while the visual input remains unchanged.

\begin{figure}[t]
    \centering
    \begin{tikzpicture}
        \begin{axis}[
            font=\small,
            width=\columnwidth,
            enlargelimits=false,
            height=6cm,
            xlabel={SNR (dB)},
            xlabel style={at={(0.5,-0.075)},anchor=north},
            ylabel={WER (\%)},
            ylabel style={at={(-0.075,0.5)},anchor=south},
            xmin=-5, xmax=20,
            ymin=10, ymax=130,
            xtick={-5,0,5,10,20},
            ytick={20,40,60,80,100,120},
            legend style={font=\small, at={(0.675,0.95)},anchor=north,legend columns=1},
            ymajorgrids=true,
            grid style=dashed,
            line width=1pt,
            mark size=1.5pt
        ]
        \addplot[color=blue,mark=*,mark size=2pt,line width=1.2pt] coordinates {
            (20,21.4) (10,29) (5,40.3) (0,56.0) (-5,78)
        };
        \addlegendentry{Our model (AV)}

        \addplot[color=green!50!black,mark=square*,dashed,mark options={solid},line width=1pt] coordinates {
            (20,32.0) (10,35.44) (5,46.04) (0,83.61) (-5,123.71)
        };
        \addlegendentry{Whisper-large}

        \addplot[color=purple,mark=diamond*,dashed,mark options={solid},line width=1pt] coordinates {
            (20,19) (10,22.38) (5,32.75) (0,59.44) (-5,113.76)
        };
        \addlegendentry{Whisper-large-v3}

        \end{axis}
    \end{tikzpicture}
    \caption{WER (\%) comparison under additive white Gaussian noise at different SNR levels.}
    \label{fig:noise_robustness}
\end{figure}

As shown in Figure~\ref{fig:noise_robustness}, despite being trained only with synthetic visual cues, our AV model exhibits greater robustness to acoustic degradation than the audio-only baselines. While Whisper-large-v3 slightly outperforms our model under clean conditions, its performance deteriorates rapidly as noise increases. In contrast, our AV model shows a more gradual degradation, outperforming both Whisper variants at low SNRs and exhibiting a flatter slope of the WER curve.

These results indicate that the synthetic visual streams can provide complementary information that remains useful as the acoustic signal becomes increasingly degraded, enabling the AVSR model to retain substantially better recognition performance under challenging acoustic conditions.

\subsection{Analysis of Synthetic Visual Speech}
\label{sec:analysis}

Having demonstrated that synthetic video improves AVSR performance, we now investigate which properties of synthetic video may explain this improvement. We examine three properties of the synthetic visual streams: whether temporal synchronization between audio and visual signals is preserved (Section~\ref{sec:coupling}), how their motion dynamics compare to real video (Section~\ref{sec:motion}), and whether mouth movement remains correlated with acoustic energy (Section~\ref{sec:energy_correlation}).

As a real-data reference, we use LIP-RTVE~\cite{lleida2019albayzin, liprtve2022lrec}, a $\approx$13h Spanish AV corpus built from broadcast television with carefully curated segmentation and alignment. Its manually reviewed recordings provide a reference for natural audiovisual coupling against which synthetic corpora can be assessed.

\subsubsection{Audiovisual Temporal Coupling}
\label{sec:coupling}

In audiovisual speech recognition, acoustic and visual information must be effectively integrated to achieve robust recognition. Unlike visual-only systems, AVSR relies on learning joint representations from synchronized audio and visual streams \cite{shi2022avsr,shi2022avhubert}. Thus, synthetic visual data that improve lipreading performance may still be unsuitable for AVSR if they introduce domain shift or temporal misalignment, potentially disrupting cross-modal learning. We therefore investigate whether our generated synthetic videos maintain genuine temporal synchronization with the corresponding audio, enabling the model to learn meaningful audiovisual correlations.

To this end, we derive a visual feature from facial landmarks and compare it against the acoustic signal on a frame-by-frame basis. For each video $v$, facial landmark detection provides, at every frame $t$, the 2D position $\mathbf{p}_i(t) \in \mathbb{R}^2$ of landmark $i \in \{1,\dots,68\}$ (\texttt{dlib} 68-point convention \cite{king2009dlib}, see Figure~\ref{fig:landmark_template}). Mouth opening $o(t)$ is defined as the mean inner-lip vertical distance, following standard landmark-based articulatory measures,
\begin{equation}
o(t) = \frac{1}{2}\Big(\lVert \mathbf{p}_{62}(t) - \mathbf{p}_{66}(t)\rVert + \lVert \mathbf{p}_{63}(t) - \mathbf{p}_{65}(t)\rVert\Big).
\label{eq:opening_raw}
\end{equation}
Since this raw distance depends on camera distance, resolution, and face size, we normalize it, following standard practice, by a per-frame face-scale reference $s(t)$, computed as the median distance across a small set of symmetric facial landmark pairs $\mathcal{P}$:
\begin{equation}
\hat{o}(t) = \frac{o(t)}{s(t)}, \qquad s(t) = \operatorname*{median}_{(i,j)\in\mathcal{P}} \lVert \mathbf{p}_i(t) - \mathbf{p}_j(t)\rVert,
\label{eq:opening_norm}
\end{equation}
where $\hat{o}(t)$ denotes the face-scale-normalized mouth opening. The set $\mathcal{P} = \{(1,15), (2,14), (3,13)\}$ (\texttt{dlib} convention, see Figure~\ref{fig:landmark_template}) corresponds to symmetric cheek landmark pairs, chosen because they remain approximately rigid during articulation and therefore provide a stable scale reference, unlike mouth-region landmarks whose relative position changes with speech. We use the median rather than the mean, as it is more robust to outlier frames arising from landmark detection errors.

The resulting signal is smoothed with a second-order Savitzky--Golay filter (7-frame window) \cite{savitzky1964smoothing} to reduce landmark-tracking jitter and standardized with a robust z-score based on the median and median absolute deviation (MAD),
\begin{equation}
v_o(t) = \frac{\hat{o}(t) - \operatorname{median}(\hat{o})}{1.4826 \cdot \operatorname{MAD}(\hat{o})}.
\label{eq:zscore}
\end{equation}
where $v_o(t)$ represents the standardized mouth-opening position. 

Having derived this visual signal, we now define its acoustic counterpart so that both can be directly compared. The acoustic signal $a(t)$ is the short-time energy (RMS envelope), computed over 25\,ms windows with 10\,ms hop. Because visual articulation is known to precede the corresponding acoustic signal due to \textit{anticipatory coarticulation}~\cite{whalen1990coarticulation, jordan1997serial}, speakers begin shaping their mouth for an upcoming sound before that sound is acoustically realized. In natural audiovisual speech, mouth movements have been reported to precede the acoustic signal by approximately 100--300\,ms~\cite{chandrasekaran2009natural}, and studies of audio-visual synchronization commonly assess correlation across a range of temporal offsets rather than at a fixed lag~\cite{slaney2000facesync,bredin2007audiovisual,sargin2007audiovisual,chung2017out}. We adopt the same approach, computing, for each lag $\tau$ in $[-\Delta, \Delta]$ with $\Delta = 1000\,$ms, the Spearman rank correlation,
\begin{equation}
\rho(\tau) = \rho_S\big(a(t-\tau),\, v_o(t)\big),
\label{eq:lag_profile}
\end{equation}
and extracting the best-lag correlation,
\begin{equation}
\rho^{*} = \max_{\tau \in [-\Delta,\Delta]} \rho(\tau), \qquad \tau^{*} = \operatorname*{arg\,max}_{\tau \in [-\Delta,\Delta]} \rho(\tau).
\label{eq:best_lag}
\end{equation}
where $\rho_S$ denotes the Spearman rank correlation coefficient between its two arguments.

To rule out correlation arising purely from the signals' own smoothness rather than genuine coupling, we build a per-video null distribution, adapting the shift-predictor logic of~\cite{perkel1967neuronal} to continuous audiovisual signals. For each video, the audio track is circularly shifted by $K=50$ random offsets of at least 1200\,ms (chosen to exceed the $\pm 1000\,$ms lag window in Eq.~\ref{eq:best_lag} to avoid overlap with the original alignment). For each shifted version, $\rho(\tau)$ is recomputed following Eq.~\ref{eq:lag_profile}, and the median best-lag correlation across shifts is taken as the null $\tilde\rho_0$.
We define the synchrony excess as
\begin{equation}
	S = \rho^{*} - \tilde\rho_0,
	\label{eq:synchrony_excess}
\end{equation}
where a high $S$ indicates genuine audiovisual coupling beyond chance level.


\begin{table}[t]
\centering
\caption{Audiovisual temporal coupling between mouth opening and acoustic energy. $\rho_0$, zero-lag correlation; $\rho^{*}$, best-lag correlation; $\tau^{*}$, lag at peak correlation; $S$, synchrony excess (correlation beyond chance). All values are dataset medians.}
\label{tab:synchrony}
\begin{tabular*}{\columnwidth}{@{\extracolsep{\fill}}lcccc@{}}
\toprule
Dataset & $\rho_0$ & $\rho^{*}$ & $\tau^{*}$ (ms) & $S$ \\
\midrule
LIP-RTVE (real)       & $-0.024$ & $0.288$ & $-160$ & $0.234$ \\
TV3Parla (synth.)     & $0.007$  & $0.314$ & $-120$ & $0.320$ \\
ParlamentParla (synth.) & $-0.050$ & $0.280$ & $-120$ & $0.258$ \\
\bottomrule
\end{tabular*}
\end{table}

Table~\ref{tab:synchrony} reports these quantities on real (LIP-RTVE) and synthetic (TV3Parla, ParlamentParla) datasets. Both real and synthetic videos exhibit strong synchrony excess ($S \approx 0.23$--$0.32$), well beyond chance level, and the optimal lags are negative in all cases ($\tau^{*} \approx -120$ to $-160$\,ms), indicating that visual motion precedes the acoustic envelope. This ordering is consistent with the 100--300\,ms anticipatory coarticulation window noted above, showing that our synthetically generated videos reproduce the physiological timing observed in natural audiovisual communication.

Given that we use the full datasets for our experiments, the large sample sizes involved may cause standard significance tests to yield statistically detectable differences even for negligible effects. We therefore use Cliff's delta ($\delta$)~\cite{romano2006appropriate}, a sample-size-independent effect size measure, to quantify the magnitude of differences in synchrony excess. Following common conventions for this statistic, we interpret $|\delta|<0.147$ as negligible, $|\delta|<0.33$ as small, $|\delta|<0.474$ as medium, and $|\delta|\geq0.474$ as large.

Table~\ref{tab:cliffs_delta} shows that differences in synchrony excess between real and synthetic datasets range from negligible ($|\delta|=0.135$ for ParlamentParla) to small ($|\delta|=0.326$ for TV3Parla). Real and synthetic videos therefore exhibit comparable levels of audiovisual coupling, indicating that synthetic mouth movements preserve the temporal structure of natural speech.

\begin{table}[t]
\centering
\caption{Effect sizes ($\delta$) for real-vs-synthetic differences in synchrony excess. Negligible effect sizes indicate comparable audiovisual coupling.}
\label{tab:cliffs_delta}
\begin{tabular*}{\columnwidth}{@{\extracolsep{\fill}}lcc@{}}
\toprule
Comparison & $\delta$ & Effect \\
\midrule
Real vs. TV3Parla       & $-0.326$ & Small \\
Real vs. ParlamentParla & $-0.135$ & Negligible \\
\bottomrule
\end{tabular*}
\end{table}

\subsubsection{Fine-Grained Visual Motion Dynamics}
\label{sec:motion}

While Section~\ref{sec:coupling} shows that the overall timing of mouth movement is preserved, this does not guarantee that the underlying motion resembles natural articulation, as synthetic generation can still introduce artifacts at a finer temporal grain~\cite{shan2022speech,liu2023synthvsr}. To complement the landmark-based opening measure (Eq.~\ref{eq:opening_norm}), we additionally quantify local pixel motion within the mouth region using optical flow.

Optical-flow magnitude is computed with the Farneb\"ack method~\cite{farneback2003two} between consecutive frames, averaged within a fixed per-video mouth region of interest (ROI). The ROI is defined as the union of per-frame landmark-derived boxes, expanded by a 20-pixel margin to avoid frame-to-frame ROI jitter. 


As shown in Table~\ref{tab:motion_velocity}, synthetic videos exhibit lower frame-to-frame changes in mouth opening than real videos ($|\Delta \hat o(t)|$, Eq.~\ref{eq:opening_norm}), with a large effect size. This pattern is consistent across both synthetic datasets ($|\delta|<0.03$ between TV3Parla and ParlamentParla), suggesting that it is primarily associated with the generation process rather than with the source corpus. In contrast, optical-flow magnitude within the mouth region remains comparable between real and synthetic videos, with only a medium effect size (Table~\ref{tab:motion_velocity}). Thus, while landmark-based articulatory dynamics are more constrained in synthetic video, local pixel motion within the mouth region remains comparatively close between synthetic and real video.

\begin{table}[t]
\centering
\caption{Articulatory motion: frame-to-frame landmark opening change and optical-flow magnitude in the mouth region (medians). Cliff's $\delta$ corresponds to the pooled synthetic-versus-real comparison.}
\label{tab:motion_velocity}
\begin{tabular*}{\columnwidth}{@{\extracolsep{\fill}}lcc@{}}
\toprule
Dataset & $|\Delta \hat{o}(t)|$ & Optical Flow \\\midrule
LIP-RTVE (real)         & 0.010 & 0.352 \\
TV3Parla (synth.)       & 0.007 & 0.516 \\
ParlamentParla (synth.) & 0.007 & 0.512 \\
\midrule
Synthetic vs. Real      & $\delta=-0.601$ & $\delta=0.447$ \\
Effect                  & Large &   Medium \\
\bottomrule
\end{tabular*}
\end{table}

This difference in landmark motion does not necessarily imply a loss of speech-relevant information. Real recordings naturally contain head-pose changes and other non-articulatory facial movements, which can introduce additional frame-to-frame variation into the mouth-opening trace, whereas our synthetic talking heads are generated from a still image with motion confined entirely to the mouth region. The narrower landmark trajectories in synthetic video likely reflect this more controlled setting rather than reduced articulatory content.

\subsubsection{Audiovisual Energy Correlation}
\label{sec:energy_correlation}

The previous analyses show that synthetic mouth movements preserve realistic timing relative to speech, but move within a narrower range than real video. What matters, however, is whether this motion still carries speech-relevant content, which we examine directly in this section. Specifically, we test whether mouth movement remains co-modulated with speech energy~\cite{grant2000visible}, that is, whether variation in one signal is systematically related to variation in the other over time. Prior work has shown that such co-modulation indicates genuine audiovisual correspondence, rather than arising incidentally from two smoothly varying, positively-valued signals~\cite{chandrasekaran2009natural}.

Chandrasekaran et al.~\cite{chandrasekaran2009natural} demonstrated this distinction using a shuffled-correlation test. Mouth area correlated with the matched acoustic envelope was compared against mismatched envelopes from other utterances by the same speaker, and correctly paired recordings yielded substantially higher correlation than mismatched pairings, which dropped to chance level. We adopt an analogous logic here, but quantify co-modulation directly as the mutual information (MI) between mouth opening $\hat o(t)$ and the acoustic energy envelope $a(t)$, estimated with the Kraskov--St\"ogbauer--Grassberger $k$-nearest neighbor ($k$-NN) estimator~\cite{kraskov2004estimating} using $k=5$ neighbors,
\begin{equation}
	\hat{I}(X;Y) = \psi(k) - \big\langle \psi(n_x+1) + \psi(n_y+1) \big\rangle + \psi(N),
	\label{eq:ksg_mi}
\end{equation}
where $\psi$ is the digamma function, $N$ is the sample size, and $n_x$, $n_y$ are the counts of neighboring points within each marginal at the $k$-NN radius. We denote the resulting estimate as $\hat{I}$ to distinguish it from the underlying mutual information. We use MI rather than the Pearson correlation used in~\cite{chandrasekaran2009natural}, since MI also captures nonlinear statistical dependencies that Pearson correlation cannot~\cite{zhao2020mutual}. Datasets are compared using Cliff's delta, as in previous sections.

As shown in Table~\ref{tab:energy_mi}, both synthetic datasets preserve audio-energy co-modulation at levels closely matching real video, with negligible effect sizes in both comparisons. This preservation holds despite the lower frame-to-frame variation in synthetic landmark motion, indicating that mouth movement remains systematically coupled to speech intensity even though its dynamics differ from real video.

\begin{table}[t]
\centering
\caption{Audiovisual energy correlation (mutual information estimate between mouth opening and acoustic energy envelope, $k=5$).}
\label{tab:energy_mi}
\begin{tabular*}{\columnwidth}{@{\extracolsep{\fill}}lcc@{}}
\toprule
Dataset & $\hat{I}(\hat{o};a)$ & Effect vs. Real \\
\midrule
LIP-RTVE (real)    & 0.0563 & Baseline \\
TV3Parla (synth.)       & 0.0574 & $\delta=-0.012$ (negligible) \\
ParlamentParla (synth.) & 0.0465 & $\delta=-0.067$ (negligible) \\
\bottomrule
\end{tabular*}
\end{table}

Taken together with the temporal coupling results in Section~\ref{sec:coupling}, these findings show that the synthetic audiovisual coupling reproduces the non-chance-level structure that characterizes genuine audiovisual speech, rather than arising as a trivial byproduct of simultaneously-generated signals. 

Because this coupling is a property of the visual signal itself, it remains unaffected by noise added to the audio at test time, offering a plausible explanation for why the visual modality continues to provide complementary, speech-related information as acoustic quality degrades (Section~\ref{subsec:noise}).

\section{Discussion}
\label{sec:discussion}

The central finding of this work is that an AVSR system can be bootstrapped for a genuinely zero-AV-resource language using synthetic visual streams as the sole source of visual supervision. The AV model trained exclusively on synthetically-generated video outperforms an identically trained audio-only model by a relative WER reduction of 15.2\% (Section~\ref{subsec:cat_model_performance}), approaches the performance of Whisper-large-v3 despite orders of magnitude less training data and roughly $5\times$ fewer parameters, and clearly outperforms Whisper-large (Section~\ref{subsec:baseline_comparison}). Notably, the multimodal advantage is not confined to clean conditions and it widens as acoustic quality degrades. Our AV model shows a markedly flatter WER curve than both Whisper baselines under additive noise (Section~\ref{subsec:noise}). These results suggest that synthetic visual streams, despite being insufficient alone for reliable transcription (Section~\ref{subsec:cat_model_performance}), provide a source of information the model can draw on.

The analyses in Section~\ref{sec:analysis} offer a plausible explanation for this pattern. The temporal relationship between mouth opening and the acoustic envelope in our synthetic corpora is well above chance and occurs at negative lags of $-120$ to $-160$\,ms (Table~\ref{tab:synchrony}), consistent with the 100--300,ms anticipatory coarticulation window reported for natural audiovisual speech~\cite{chandrasekaran2009natural}. Our synthetically generated videos reproduce not just loose synchronization, but the physiologically correct ordering of visual and acoustic events. The strength of this coupling remains comparable to real video, with effect sizes ranging from negligible to small (Table~\ref{tab:cliffs_delta}).

Optical-flow magnitude within the mouth region is also comparable between synthetic and real video, despite narrower frame-to-frame motion in synthetic landmark trajectories (Table~\ref{tab:motion_velocity}), which is expected given the controlled nature of talking-head generation. Most directly relevant to the AVSR results, the mutual information between mouth opening and acoustic energy is close to the real-video estimate for both synthetic datasets (Table~\ref{tab:energy_mi}). Since this coupling collapses under non-corresponding audiovisual pairing (Section~\ref{sec:energy_correlation}), its preservation here points to a genuine, speech-specific dependency rather than an artifact of two smoothly varying signals.

These properties explain both halves of the empirical result. Under clean conditions, a correctly timed envelope-correlated cue supplies a representation of syllabic and prosodic structure that the model can exploit more readily than the acoustic signal alone, plausibly accounting for the WER gain in Section~\ref{subsec:cat_model_performance}. Under acoustic degradation, this visual cue is by construction unaffected by noise added to the test-time audio channel, which may explain why the AV model's advantage over audio-only baselines widens as SNR decreases (Section~\ref{subsec:noise}). We hypothesize that AV-HuBERT's modality-dropout pretraining~\cite{shi2022avhubert} is particularly well suited to exploiting this signal, as it gives the encoder a robust, noise-independent proxy for acoustic energy to fall back on when the acoustic stream is uninformative.

These results also extend our earlier findings~\cite{buitrago2026improving}. Our prior work showed that synthetic supervision can provide additional value when combined with an existing real AV corpus. The present study shows that synthetic visual data alone can also provide sufficient speech-correlated and temporally aligned visual information to train a competitive AVSR system without any real visual training data. This finding is particularly relevant for under-resourced languages, where the availability of paired audiovisual data remains the main bottleneck.

\section{Conclusion}

We have shown that purely synthetic video, generated via lip synchronization on static facial images, can provide effective articulatory cues for AVSR in a zero-AV-resource setting. Using over 700 hours of synthetic Catalan talking-head videos, we trained an AVSR model that outperforms an identically trained audio-only baseline and retains the benefits of multimodal integration under acoustic degradation. Despite being substantially smaller and trained on far less data, the model achieves performance competitive with state-of-the-art ASR baselines~\cite{radford2022whisper} under clean conditions, and outperforms them under noisy conditions.

These findings demonstrate that synthetic video derived solely from real audio could serve as an effective source of visual supervision without requiring native audiovisual training data. Together with the extended semi-automatic annotation pipeline, which facilitates the creation of real audiovisual resources, our approach provides a scalable path toward AVSR for under-resourced languages.

\section*{Declarations}

\bmhead{\small Abbreviations}
A - Audio-only modality \\
ASR - Automatic Speech Recognition \\
AV - Audiovisual \\
AV-HuBERT - Audio-Visual Hidden Unit BERT \\
AVSR - Audiovisual Speech Recognition \\
FFHQ - Flickr-Faces-HQ dataset \\
GAN - Generative Adversarial Network \\
GUI - Graphical User Interface \\
k-NN - k-Nearest Neighbor \\
MAD - Median Absolute Deviation \\
MI - Mutual Information \\
RMS - Root Mean Square \\
ROI - Region of Interest \\
SNR - Signal-to-Noise Ratio \\
SOTA - State-of-the-Art \\
V - Video-only modality \\
WER - Word Error Rate \\

\bmhead{\small Availability of data and materials}

The publicly available datasets used in this study can be obtained from the original sources cited throughout the manuscript. The implementation of the proposed methodology, together with the scripts required to reproduce the experiments, is publicly available at \url{https://github.com/Pol-Buitrago/SynthAVSR}. 

The synthetic audiovisual datasets and the manually annotated Catalan audiovisual benchmark generated during this study are not publicly available because they are derived from third-party audiovisual content that cannot be redistributed. Additional information may be provided by the corresponding author upon request, subject to the licensing restrictions of the original data sources.

\bmhead{\small Competing interests}
The authors declare that they have no competing interests.

\bmhead{\small Funding}
Not applicable.

\bmhead{\small Authors' contributions}

PB conceived the study, designed the methodology, developed the synthetic data generation pipeline, performed the experiments, conducted the formal analysis, and wrote the manuscript. JH supervised the research, and reviewed the manuscript. All authors read and approved the final manuscript.

\bmhead{\small Acknowledgements}
Not applicable.

\bigskip
\bibliography{custom}

\end{document}